\documentclass[12pt]{article}     
\usepackage{amsmath} 
\usepackage{amsfonts} 
\usepackage{amssymb}

\def\beq{\begin{equation}}
\def\eeq{\end{equation}}
\def\bea{\begin{eqnarray}}
\def\eea{\end{eqnarray}}

\newcommand{\bo}{\raise-1mm\hbox{\Large$\Box$}}              
\def\lb{\left(}
\def\rb{\right)}

\begin{document}
\footnotetext[1]{D\'epartement de  physique, Universit\'e  de  Montr\'eal, C.P. 6128, succursale centre-ville, Montr\'eal, Qu\'ebec, Canada, H3C 3J7, e-mail: jonathan.belletete@umontreal.ca} 
\footnotetext[2]{Groupe de physique des particules, D\'epartement de  physique, Universit\'e  de  Montr\'eal, C.P. 6128, succursale centre-ville, Montr\'eal, Qu\'ebec, Canada, H3C 3J7, e-mail: paranj@lps.umontreal.ca}

\title{On negative mass}

\author{Jonathan Bellet\^ete${}^{1}$ and M. B. Paranjape${}^{2}$}

\date{March 31st, 2013 }

\maketitle
\centerline{Essay written for the Gravity Research Foundation 2013 Awards for Essays on Gravitation}

\begin{abstract} 
The Schwarzschild solution to the matter free, spherically symmetric Einstein equations has one free parameter, the mass.  But the mass can be of any sign.  What is the meaning of the negative mass solutions?  The answer to this question for the case of a pure Schwarzschild negative mass black solution is still elusive, however, in this essay, we will  consider negative mass solutions within a Schwarzschild-de Sitter geometry.  We show that there exist reasonable configurations of matter, bubbles of distributions of matter, that satisfy the dominant energy condition everywhere,  that are non-singular and well behaved everywhere, but correspond to the negative mass Schwarzschild-de Sitter geometry outside the matter distribution.  These negative mass bubbles  could  occur as the end state of a quantum tunnelling transition.  
\end{abstract}                    
\vfill\eject
\section{Negative mass particles}
It has been the common understanding that negative mass Schwarzschild solutions, in general relativity,  have no physical meaning.  On the other hand, it has also been realized that classical solutions, no matter how crazy, seem to have some meaning in the full quantum theory.  Take for example the case of instantons, classical solutions in imaginary time.  On the face of it, these have no physical relevance, however, they turn out to give us a powerful method to compute tunnelling amplitudes \cite{coleman}.  No initial configuration of matter, that satisfies the dominant energy condition, will collapse through classical time evolution  to a negative mass configuration \cite{penrose,hawking-ellis}.  Matter will only collapse to positive mass black hole configurations.  Perhaps the existence of the negative mass solutions is simply a sign that the classical theory is incomplete.  These solutions are only singular at one point and the singularity is naked, but away from the singularity they are very well behaved.  Perhaps their defective nature is only a problem due to our incomplete understanding of gravity.  It is believed that the full, quantum theory of gravity will have no unphysical configurations, and quite possibly the negative mass Schwarzschild solutions may have a non-trivial role in this theory, with the problem of the naked singularity resolved by  ultra-violet completeness.  It would be important to find the physical meaning of the negative mass solutions.  In this essay we find that negative mass solutions in a de Sitter geometry are perfectly physical.  
 
But if they exist, particles of negative mass are strange beasts.  Because of the equivalence principle, they are actually attracted by particles of positive mass suffering the universal acceleration $a=\frac{GM}{r^2}$.  On the other hand, negative mass particles act as repulsive sources for all other matter, indeed, the acceleration of a particle in the presence of a negative mass particle is $a=\frac{G(-M)}{r^2}$ giving rise to repulsion/anti-gravity \cite{bondi}.



The metric of the negative mass Schwarzschild metric,  a solution of the vacuum Einstein equations,  is given by
\beq
d\tau^2=\lb 1-2 G (-M)/r\rb dt^2 -\frac{dr^2}{\lb 1-2 G (-M)/r\rb}-r^2d\theta^2-r^2\sin^2\theta d\phi^2\label{1}
\eeq
with $M$ taken positive.  This solution admits a (naked) singularity at $r=0$ which is not cloaked behind an event horizon.  The singularity notwithstanding, this metric must describe some aspects of a negative mass particle.  Singularities are commonly accepted as artefacts of some idealization in the effective description that is being adopted,  of the system under consideration.  For example, point charge approximations in electrodynamics with their attendant infinite energy density and infinite forces, are of little concern when studying the low energy classical dynamics of charged particles.  We are comfortable in the understanding that at very short distances, the point charge approximation breaks down, and the charges are actually smeared over a finite spatial volume.  The same can be said of the positive mass Schwarzschild solution,  obtained by changing $-M\to M$ in the metric above, Eqn. \eqref{1}.  The exterior geometry of stars is considered to be exactly Schwarzschild, the vacuum Einstein equations being valid \cite{weinberg}.  But as soon as we hit the star's surface, there is a non-zero energy momentum tensor, and the vacuum Einstein equations are no longer applicable.  The metric is deformed so that the event horizon and the black-hole singularity are simply absent.  The content of this essay is to examine whether it is possible to do the same sort of deformation for the case of negative mass point gravitational particles.  

The Einstein equations in the presence of energy-momentum are given by
\beq
G_{\mu\nu}[g_{\lambda\rho}]=8\pi G\,\, T_{\mu\nu}\label{2}.
\eeq 
The basic idea of smearing out a singular metric corresponds to replacing  the singular metric $g_{\mu\nu}\to\tilde g_{\mu\nu}$ which has no singularities, and then using the LHS of Eqn. \eqref{2} to define what energy-momentum is required for a consistent solution.  This can be easily done for both positive and negative mass Schwarzschild solutions, the simplest deformation would be to replace $M\to M(r)$ and ensure that $M(r)\to 0$ as $r\to 0$ sufficiently fast yielding the required energy-momentum distribution for a non-singular, negative mass particle.  However, the situation is not so simple.  

Various positive energy theorems \cite{pet} state that under the assumption of asymptotic flatness and that the the dominant energy condition is satisfied, the mass parameter of a distribution of matter must always be positive. Thus negative mass configurations may not form from  the normal classical evolution of some well defined, non-singular, initial matter configuration that satisfies the dominant energy condition. Therefore, under these assumptions, any non-singular, negative mass metric must correspond to a non-singular energy-momentum tensor that somewhere violates the dominant energy condition  in its interior.  If a mass distribution violates the dominant energy condition, then the matter can be seen to be moving outside the light-cone, which is intolerable.  

However, the assumption of asymptotic flatness is crucial.   Relaxing this condition, the positive energy theorems no longer apply.   For example, an infinite class of exact although singular  solutions exist that correspond to a negative mass black hole in a de Sitter background.  The metric is given by
\beq
d\tau^2=\lb 1-\frac{(\Lambda/3) r^3-2 G M}{r}\rb dt^2 -\frac{dr^2}{\lb 1-\frac{ (\Lambda/3) r^3-2 GM}{r}\rb}-r^2d\theta^2-r^2\sin^2\theta d\phi^2\label{4}
\eeq
which is an exact  solution of  Einstein's equations in the presence of a cosmological constant  
\beq
G_{\mu\nu}[g_{\lambda\rho}]= \Lambda g_{\mu\nu}.\label{3}
\eeq 
 The RHS of Eqn. \eqref{3}  can be interpreted as the energy-momentum tensor of a ``dark-energy'' type of matter $T_{\mu\nu}=(\Lambda/8\pi G) g_{\mu\nu}$.     We will show that the metric \eqref{4} can be deformed to a completely non-singular metric that everywhere respects the dominant energy condition.

\begin{section}{Dominant energy condition}
The metric for a spherically symmetric, static space-time in Schwarzschild coordinates is given by
\begin{equation}
d\tau^2 = \left(1-2\frac{ M(r)}{r}\right)dt^2 - \frac{1}{\lb 1-2\frac{M(r)}{r}\rb} dr^2 - r^2d\theta^2 - r^2 \sin^2(\theta)d\phi^2
\end{equation}
where $M(r)$ is the effective mass of the space-time.  We have set $8 \pi G$  and the speed of light equal to 1.  The corresponding energy-momentum tensor is
\begin{equation}
T^{0}_{0} = T^{1}_{1} = \frac{2 M'(r)}{r^2},\quad
\quad
T^{2}_{2} = T^{3}_{3} =  \frac{M''(r)}{r}.
\end{equation}
 The dominant energy conditions states that for all timelike or lightlike vector $u$, $T^{0 \nu}u_{\nu} \geq 0$ and  $T^{\mu \nu}u_{\nu} T_{\mu \alpha}u^{\alpha}  \geq 0$. Choosing a Lorentz frame, a general timelike vector take the form $u^{\mu} = \frac{1}{\left( 1- a^2 - b^2 - c^2\right)^{1/2}}{ (1, a, b, c)}$ with $ 1- a^2 - b^2 - c^2>0$. The first inequality implies
\begin{equation}
T^{0 \nu}u_{\nu} =  \frac{2 M'(r)}{r^2 \sqrt{1-a^2-b^2-c^2}}\geq 0
\end{equation}
which is equivalent to
\begin{equation}
M'(r) \geq 0 \label{DC1}.
\end{equation}
The second inequality  implies
\begin{equation}
T^{\mu \nu}u_{\nu} T_{\mu \alpha}u^{\alpha} =  \left( \frac{4 M'(r)^2}{r^4}-\frac{\left(b^2+c^2\right) \left(r^2 M''(r)^2-4
   M'(r)^2\right)}{\left(1-a^2-b^2-c^2\right) r^4}\right) \geq 0.
\end{equation}
Since $1-a^2-b^2-c^2$ can be arbitrarily small, this is true for all timelike vectors only if
\begin{equation}
\left(r M''(r)-2 M'(r)\right) \left(r M''(r)+2 M'(r)\right)\leq 0.
\end{equation}
Using (\ref{DC1}), this is equivalent to
\begin{equation}
\frac{d}{dr}\left(\frac{M'(r)}{r^2} \right) \leq 0 \label{EC1}
\end{equation}
\begin{equation}
\frac{d}{dr}\left(M'(r)r^2 \right) \geq 0 \label{EC2}
\end{equation}
The dominant energy condition is thus equivalent to these two inequalities.

\end{section}
\begin{section}{Particular class of solution}
We consider specific solutions corresponding to a bubble which has a non-singular core, a transition shell and an exterior which is an exact negative mass Schwarzschild-de Sitter solution.  For if $r<R_1$ we take 
\begin{equation}
M(r) = m + \frac{\lambda  r^3}{6 } - \frac{q}{2 r}.
\end{equation}
(Granted this function is singular at $r=0$, but it can be easily smoothed out by replacing $q\to q(r)$ near $r=0$. This singularity has no bearing on the dominant energy condition.)   For $r>R_2>R_1$, we take
\begin{equation}
M(r) = M + \frac{\Lambda r^3}{6}
\end{equation}
where we will show that $M$ can be taken negative and for $R_1 \leq r \leq R_2$, we find the interpolation:
\begin{multline}
M(r)=-\frac{2 q \int_{R_1}^r z^3 \int_{R_1}^z J_{n p}(\xi ) \,
   d\xi  \, dz}{R_1^5 J_{n p}\left(R_1\right)}+\left(m-\frac{q}{2
   R_1}+\frac{\lambda  
   R_1^3}{6}\right)\\
   +\frac{\left(r^3-R_1^3\right)}{3 R_1^2} \left(\frac{q}{2
   R_1^2}+\frac{\lambda   R_1^2}{2}\right)
\end{multline}
with
\begin{equation}
J_{n p}(r) = \frac{1}{r^5}\int_r^{R_2} \left(z-R_1\right){}^n \left(R_2-z\right){}^p
   \, dz
\end{equation}
with $p$ and $n$ arbitrary positive integers.
Furthermore, 
\begin{equation}
M\equiv M(R_2)  = m-\frac{2 q}{3} \frac{ \int_{R_1}^{R_2} \frac{1}{z}\left(z-R_1\right){}^n \left(R_2-z\right){}^p
   \, dz}{ \int_{R_1}^{R_2} \left(z-R_1\right){}^n \left(R_2-z\right){}^p \, dz}
\end{equation}
\begin{equation}
\lambda  =\Lambda -\frac{q}{R_1^4} \left(1-\frac{4 \int_{R_1}^{R_2} J_{n
   p}(\xi ) \, d\xi }{R_1 J_{n p}\left(R_1\right)}\right).
\end{equation}
This configuration represents a bubble with an interior radius  $R_1>0$ containing a charged black hole of mass $m>0$ and charge $q>0$ and filled with a cosmological constant $\lambda >0$. The exterior radius of the bubble is $R_2$ and behaves like a black hole of mass $M<m$ with a cosmological constant $\Lambda>0$. A little work shows that the metric, the connexion and the Einstein tensor produced by this choice of $M(r)$ are all continuous. Furthermore, it can be easily verified that the dominant energy condition is everywhere satisfied.   Also, since $M<m$, for all $p$ and $n$, there will always be a value of $q$ such that $M<0$ while still having $m>0$. The exterior of this bubble would thus appear as a part of de Sitter space-time with a negative-mass black hole in it.
\end{section}
\begin{section}{Conclusion}
Negative mass configurations within a de Sitter universe context must have important consequences in the early universe.  In the inflationary epoch \cite{guth}, the universe is essentially a de Sitter space-time. Pair production of negative-positive mass pairs would give rise to a strange, neutral, gravitational plasma, which could dominate the behaviour of gravitational waves in the early universe.  Indeed, in contradistinction to oppositely charged pairs of particles which attract one another, positive and negative mass pairs of particles chase after each other.  The negative mass is attracted by the positive mass, but the positive mass is repelled by the negative mass.  Thus as the negative mass approaches the positive mass, it in turn moves away due to the repulsion.  Thus they chase after one another, the negative mass chasing the positive mass.    Such behaviour is not in contradiction to Newton's law of conservation of momentum.  Indeed the positive mass particle has positive momentum  while the negative mass particle has negative momentum, as its mass is negative, yielding that the total momentum zero.  Such a plasma seems to be fundamentally unstable.  However, for an infinite medium, such as the universe, there is no problem.  Any pair escaping from a given space-time domain will be replaced  on average with one entering from without.  Overall mass neutrality is clearly maintained.   Thus a gravitational plasma of positive and negative mass particles will not disappear.   One would imagine that the gravitational plasma would screen all low frequency gravitational waves.  Therefore, the low frequency gravitational content of the early universe will  be fundamentally prohibited from our view. 

\end{section}
\section*{Acknowledgements}
We thank NSERC of Canada and FQRNT of QuŽbec for financial support.

\end{document}